\documentclass[11pt]{article}
\usepackage{acl}
\usepackage{times}
\usepackage{latexsym}
\usepackage[T1]{fontenc}
\usepackage[utf8]{inputenc}
\usepackage{microtype}
\usepackage{graphicx}
\usepackage{booktabs}
\usepackage{array}
\usepackage{amsmath,amssymb}
\usepackage{xcolor}
\usepackage{multirow}
\usepackage{subcaption}
\usepackage{pifont}
\usepackage{enumitem}
\usepackage{xurl}
\newcommand{\skill}[1]{\texttt{skill@#1}}
\newcommand{\none}{\texttt{none}}

\newcommand{\bfcl}{\textsc{bfcl}}
\newcommand{\mathbench}{\textsc{math}}

\title{Attributing Structured-Output Gains in Function Calling:\\Interface Alignment versus Procedural Transfer}

\author{
Wanyi Chen$^{1,2}$, Daoyuan Chen$^{3}$, and Fang Kong$^{1,2*}$ \\
$^{1}$School of Computer Science and Technology, Soochow University \\
$^{2}$Jiangsu Key Lab of Language Computing, Suzhou 215123, China \\
$^{3}$Tongyi Lab, Alibaba Group \\
\small{\texttt{berryccc0904@gmail.com, daoyuanchen.cdy@alibaba-inc.com, kongfang@suda.edu.cn}} \\
\small{$^{*}$Corresponding author}
}

\begin{document}
\maketitle

\begin{abstract}

Function-calling benchmarks assess whether language models can select
appropriate tools and supply the required arguments. Because their scores also
depend on expressing calls in an evaluator-specific format, gains from
prompt-prepended skills may reflect interface compliance rather than
transferable task-solving procedures.
We introduce a four-layer
gain-attribution protocol for prompt-prepended skill injection, combining
canonicalized rescoring, format-only controls, repaired/balanced induction, and
portability checks.
Applied to the Berkeley Function Calling Leaderboard (BFCL)
and scoped with API-Bank, MATH-500, and
MultiHop-RAG, the protocol shows that several apparent gains are better
attributed to interface alignment than to procedural transfer: format-only
prompts match or exceed full skills in key BFCL cells, repaired/balanced
induction removes the largest sub-frontier gains, and API-Bank target-native
gains are matched within 0.5 percentage points (pp) by length-matched generic
procedural prompts. These
findings treat format compliance as a useful engineering capability while
clarifying what a structured-output score certifies. We release
\textsc{bfcl-canonical} and recommend canonicalized metrics, balanced
induction, and format-only baselines for function-calling skill-gain
attribution. Code is available at
\url{https://github.com/wanyichen06/skill-injection-attribution}.

\end{abstract}

\section{Introduction}
\label{sec:intro}

Language models increasingly interact with external tools and APIs,
making function calling a central capability of tool-using agents.
Function-calling benchmarks typically evaluate whether a model selects the
correct tool, supplies the required arguments, and emits a machine-readable
call. These requirements make benchmark scores operationally useful, but they
also combine task decisions with compliance with the evaluator's output
interface.

A common way to improve tool-using agents is to prepend reusable
instructions, or ``skills,'' distilled from successful trajectories
~\citep{wang2023voyager, trace2skill2024}. A score increase after skill
injection is often interpreted as evidence that task-solving procedures have
transferred across models or tasks. In function calling, however, the same
skill may also teach the output schema preferred by the evaluator. The
resulting gain may therefore reflect better tool and argument selection, better
interface compliance, or both.

This raises a central question: \textbf{when prompt-prepended skill text
improves function-calling scores, what attribution does the evidence justify?}

A score-only answer is insufficient for this distinction. If a skill says how
to wrap a function call, the measured score may improve even when tool and
argument selection do not. Conversely, a semantically correct call can be
marked wrong when it uses a wrapper key or API syntax that the scorer does not
accept. We therefore separate two questions that are often merged: whether the
model selected the right action, and whether it expressed that action in the
evaluator's preferred interface.

Consider a simple item whose correct call is \texttt{math.gcd(a, b)}.
The model may select the right function and arguments but use
\texttt{function\_name} and \texttt{parameters}; the scorer expects
\texttt{name} and \texttt{arguments}. Strict scoring marks the output
incorrect even though the function and arguments are correct.
Canonicalized scoring isolates this wrapper-key mismatch.

\paragraph{Inconsistent evidence.}
Prior negative-transfer, metric-sensitivity, and contract-skill results
~\citep{li2026skillsbench,multi2single2024,contractskill2024} motivate direct
attribution: when benchmark scores improve, what mechanism is responsible?

\begin{figure*}[t]
    \centering
    \includegraphics[width=0.95\textwidth]{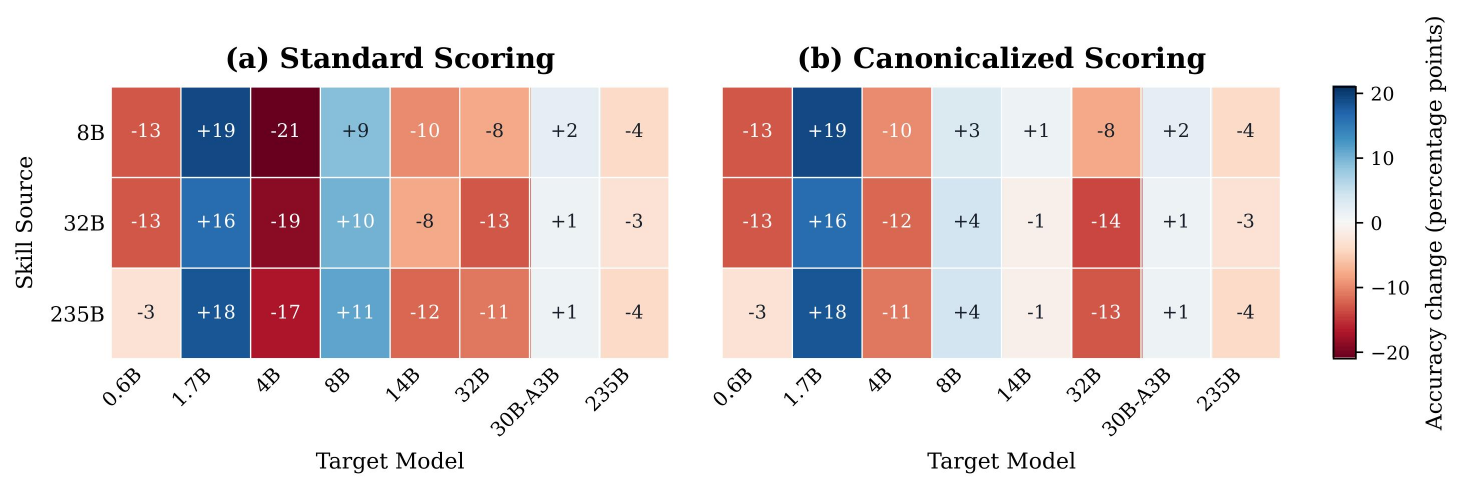}
    \caption{Effect of skill injection ($\Delta$ accuracy in percentage points (pp)) under standard vs.\ canonicalized scoring on BFCL. (a) Standard scoring shows large positive and negative effects. (b) Canonicalized scoring removes key-alias sensitivity for some cells, but many residual harms remain, showing why canonicalized metrics must be paired with mechanism controls and portability probes.}
    \label{fig:hero}
\end{figure*}

\paragraph{Function-calling skill-gain attribution.}
We propose a four-layer attribution protocol for prompt-prepended
function-calling skill gains in structured-output audits based on the Berkeley
Function Calling Leaderboard (BFCL) and API-Bank.
We instantiate the protocol primarily on BFCL, use API-Bank as a
cross-contract diagnostic, and treat MATH-500 and MultiHop-RAG as scope checks.
Our conclusions therefore concern BFCL/API-Bank-style structured-output
evaluation rather than all forms of agentic reasoning.
First, canonicalized rescoring separates wrapper-key aliasing from task-policy
changes. Second, format-only controls test whether an explicit interface
contract can reproduce the gain. Third, repaired extraction and balanced
induction test whether the effect depends on source-model style or
high-scoring demonstrations. Fourth, cross-family and cross-benchmark checks
test whether any remaining gain behaves like transferable procedure. Applying this
protocol shows, in the BFCL/API-Bank-style setting studied here, that several gains
are real under the scored interface, but their
source is often interface alignment rather than robust procedural
transfer. Figure~\ref{fig:hero} summarizes the attribution.

\paragraph{Contributions.}
This paper makes three contributions.

\begin{itemize}[leftmargin=*,itemsep=4pt]
\item We formulate \textbf{function-calling skill-gain attribution} as a
reporting problem for BFCL/API-Bank-style structured-output audits: determining
whether a prompt-prepended skill gain reflects task-policy change, interface
alignment, or both.

\item We propose and instantiate a \textbf{four-layer attribution protocol} for
function-calling skill injection, using canonicalized rescoring, format-only
controls, induction-robust extraction, and portability checks.

\item We find that, in the BFCL/API-Bank-style setting studied here, several
large gains are more consistent with \textbf{interface alignment} than robust
procedural transfer: format-only prompts can match or exceed full skills,
the largest gains fail under repaired/balanced induction controls, and matched
generic API-Bank prompts nearly reproduce extracted-skill gains. We release
\textbf{\textsc{bfcl-canonical}} and reporting guidelines for future
function-calling skill-gain claims.
\end{itemize}

\section{Related Work}
\label{sec:related}

\paragraph{Skill extraction, reuse, and negative transfer.}
Tool-use work connects language models to external actions and APIs
~\citep{yao2023react,schick2023toolformer,qin2024toolllm,patil2024gorilla},
while trajectory-derived systems store or distill reusable skills
~\citep{wang2023voyager,trace2skill2024}; agent-design and guided-replay
systems synthesize agent designs or training trajectories
~\citep{hu2024automated,xu2024agenttrek}.
Recent work shows that reuse is fragile across tasks, metrics, and contracts
~\citep{li2026skillsbench,multi2single2024,contractskill2024}. We ask whether a
text skill's structured-output gain reflects procedure or evaluator-preferred
interface conformance.

\paragraph{Structured-output benchmarks and format sensitivity.}
Function-calling benchmarks such as BFCL~\citep{patil2025bfcl},
API-Bank~\citep{li2023apibank}, \mbox{$\tau$-bench}~\citep{yao2025taubench},
and NESTFUL~\citep{basu2025nestful} score tool use through concrete output
contracts. These contracts can affect behavior apart from task content: format
restrictions can degrade performance, task solving can separate from output
formatting, and function names matter
~\citep{tam2024letme,decog2024,hammer2024}. We treat format as an attribution
mechanism through canonicalized scoring, format/wrong-alias prompts, and schema
rotation.

\paragraph{Evaluation diagnostics for skill-induced gains.}
Agent evaluation is vulnerable to small interface mismatches, and broader
evaluation work argues against treating a single score as a complete account of
model behavior~\citep{liang2023helm}. EigenData~\citep{eigendata2024} studies
function-calling data synthesis, auditing, and repair. Our contribution is
different: a claim-level attribution protocol that combines same-output
canonicalized rescoring, format-only controls, repaired/balanced induction, and
portability checks to decide when a structured-output gain justifies procedural
transfer rather than interface alignment.

\section{Evaluation Overview and Experimental Setup}
\label{sec:setup}

\begin{figure*}[t]
\centering
\includegraphics[width=0.90\textwidth]{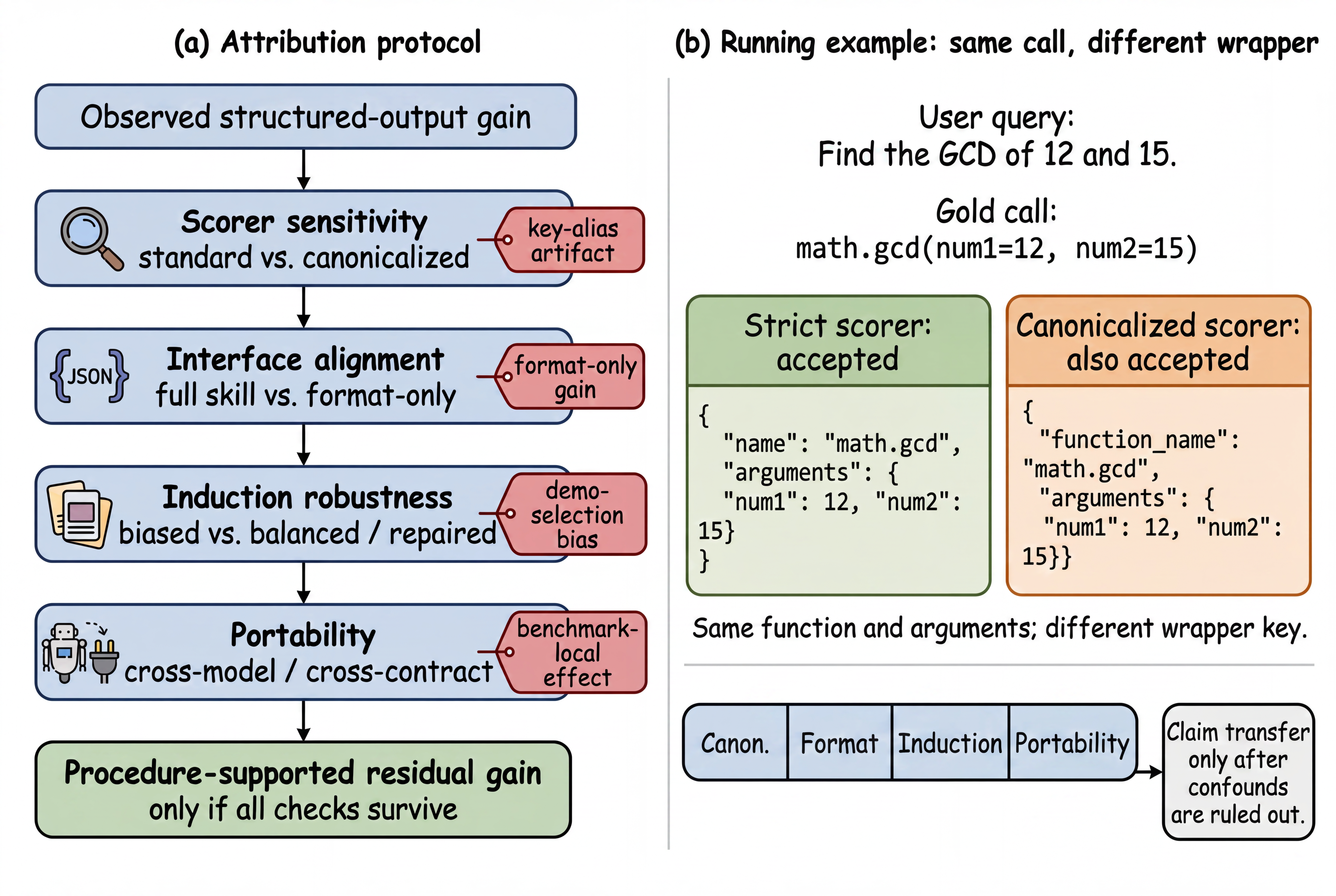}
\caption{Compact attribution protocol and key-alias running example. A gain is treated as procedural transfer only after scorer, format-only, induction, and portability controls leave a residual; Appendix~\ref{sec:case_studies} gives item-level examples.}
\label{fig:attribution_protocol}
\end{figure*}

We test whether trajectory-derived skill gains reflect procedural transfer,
interface alignment, or both using the protocol in
Figure~\ref{fig:attribution_protocol}.

The next section reports the attribution questions and corresponding tests.
Here we define only the evaluation objects, scoring variants, controls, and
scope boundaries used by those tests. We combine controlled local mechanism
audits with full
non-live official-prompt audits over the four BFCL non-live categories for
selected model/condition suites; neither is reported as a leaderboard
submission. Unless explicitly stated otherwise, BFCL-CANONICAL and
API-Bank results in this paper are controlled local diagnostics or selected
non-live audits, not official leaderboard submissions or official leaderboard
scores. BFCL-CANONICAL uses a canonicalized local scorer, while API-Bank reports
native API/required-parameter match rather than official leaderboard accuracy.

\paragraph{Primary comparisons.}
The main comparisons are full skill versus format-only on BFCL, original
versus repaired/balanced induction on the same rows, and API-Bank skill versus
length-matched generic prompts under native, strict, and row-fresh diagnostics.
Official-prompt, cross-family, prompt-variant, and case-study analyses are
scope checks; we do not average them into a headline effect.

\paragraph{From diagnostics to attribution labels.}
Paired tests and bootstrap intervals are the primary evidence when paired rows
are available; a conservative margin of $\tau=3$ percentage points is only
a descriptive screen for unpaired local matrices. Following the
terminology in Figure~\ref{fig:attribution_protocol}, we attribute a gain to
\emph{interface alignment} when canonicalization or format-only controls account
for it. We assess \emph{induction robustness} by testing whether repaired or
balanced extraction preserves the effect rather than removing it or flipping its
sign. A \emph{procedure-supported residual gain} must remain above format-only,
neutral, and generic controls after canonicalization and induction-robustness
checks; \emph{portability} then tests whether that residual keeps its direction
across models, output contracts, or benchmarks before we make a transfer claim.
These terms describe what the evidence supports; they do not identify hidden
causal mechanisms.

\paragraph{Models.}
We evaluate 8 Qwen3 models~\citep{qwen3report2025}: 0.6B--235B-A22B,
spanning a 400$\times$ parameter range. For cross-family validation
(\S\ref{sec:crossfamily}), we also evaluate Llama-3.1 (8B, 70B) and Gemma-3
(4B, 12B, 27B) on BFCL under \none{}, \skill{235B}, and
\texttt{format\_only}.

\paragraph{Benchmarks.}
We use four primary benchmarks: \textbf{\bfcl{}}~\citep{patil2025bfcl}, a 300-item
function-calling subset evaluated by AST accuracy; \textbf{\mathbench{}-500}
~\citep{lightman2023lets}, a locally filtered 470-problem subset evaluated by
sympy equivalence; \textbf{MultiHop-RAG}~\citep{tang2024multihop}, a local
300-question oracle-evidence QA subset evaluated by F1; and
\textbf{API-Bank}~\citep{li2023apibank}, locally processed Levels 1--3 with
172, 217, and 119 API-calling items.

For BFCL, the local matrix uses our fixed BFCL-v2 JSON subset and standard
all-call AST comparator to hold items, prompts, and raw outputs constant while
varying scorer canonicalization and wrapper expectations. We pair this local
mechanism audit with BFCL official-prompt non-live audits on 1{,}040 upstream
rows from the four official non-live categories
(Appendix~\ref{sec:appendix_official_controls}); these runs use the upstream
prompt-mode handler/evaluator and paired row tests, but remain non-live
selected-model audits rather than full leaderboard submissions.

\paragraph{Conditions.}
Each model-benchmark pair has up to five conditions: \none{},
\textsc{BFCL-Contract} (our BFCL-style JSON prompt, which specifies
\texttt{"name"} and \texttt{"arguments"}), and skills extracted from 8B,
32B, or 235B source models on the same benchmark. A frozen injection
template keeps all prompt text except the skill block identical.
Here, \none{} means no injected skill, \skill{235B} denotes a skill
extracted from high-scoring Qwen3-235B-A22B trajectories, and \texttt{format-only}
denotes the required output contract without extracted task-solving procedure.
Auxiliary controls are grouped by the attribution question they test, not by
whether they favor the hypothesis: format-only and alias-only controls test
interface alignment; content-only, repaired, and balanced-induction
controls test residual procedural content; matched neutral and generic
procedural prompts test whether extracted wording adds value beyond length- and
contract-matched instructions. We interpret controls symmetrically: matching
the skill weakens extraction-specific attribution, while harmful or
model-dependent behavior is reported as a boundary rather than discarded.

For attribution, we track three observable components of a scored call:
tool identity, required argument content, and wrapper/interface form. These
categories describe the evaluated output; we do not assume that the underlying
model abilities are independent. Canonicalization changes only the scorer,
format-only supplies an interface contract without extracted procedure, and
repaired/balanced or cross-contract controls vary the induction or evaluation
setting. These controls are therefore stress tests, not independent latent-module
interventions. In particular, format-only is a sufficiency control: matching a
full skill does not prove that useful procedural behavior is absent.

\paragraph{Skill extraction.}
\emph{Phase 1 (original) skills} were extracted per benchmark by GPT-4o
from high-scoring \none{} demonstrations. \emph{Phase 2 (repaired) skills}
were re-extracted by Qwen3-32B from balanced high/low-scoring subsets.
The target, BFCL rows, injection
position, and scorer remain fixed, while the extractor family, prompt, and
demonstration pool change. Thus, repaired/balanced induction is a sensitivity
test for the induction design rather than a single-factor causal intervention.
Phase~1 is the historical input being tested; repaired, balanced,
extractor-by-demonstration, and
official-prompt controls jointly test whether attribution survives after the
original high-score demo selection and extractor style are no longer fixed.
Appendix~\ref{sec:appendix_repro_checklist} summarizes the retained artifacts
for regenerating tables and paired statistics. We use ``repaired'' for
the Phase~2 re-extraction, ``balanced'' for high/low-scoring demonstration pools,
and ``extractor-by-demonstration'' for comparisons that vary these induction
factors.

\paragraph{Evaluation protocol.}
For \bfcl{}, we compute three scoring variants.
Let $\hat{y}_i$ denote the model's predicted function call for item $i$,
and let $\kappa(\hat{y}_i)$ extract the JSON key used for the function name
(e.g., \texttt{"name"}, \texttt{"function"}, or \texttt{"function\_name"}).
Define the canonical projection $\phi$ that maps all key variants to a
single canonical form while preserving argument content:
\begin{equation}
  \label{eq:canonical}
  \mathrm{Acc}_{\mathrm{can}} = \frac{1}{n}\sum_{i=1}^{n}
    \mathbf{1}\!\left[\mathrm{AST}\!\left(\phi(\hat{y}_i),\; y_i\right) = 1\right]
\end{equation}
where $\mathrm{AST}(\cdot,\cdot)$ is the BFCL all-call AST comparator and $y_i$ is the ground truth.
Standard accuracy uses $\hat{y}_i$ directly (without $\phi$);
format-normalized accuracy restricts the sum to items where
$\kappa(\hat{y}_i) = \texttt{"name"}$.

For schema rotation on Qwen3-8B and 14B, minimal format cues request three
equivalent wrappers; we score outputs canonically and under strict
schema-specific expectations. The BFCL schema-control ladder additionally
compares schema-only, schema-plus-example, and length-matched neutral
instructions on Qwen3-1.7B, 8B, 14B, and 32B.

Canonicalization is a narrow de-aliasing diagnostic, not a semantic
``forgiving'' evaluator. It merges only wrapper-key variants that preserve the
same semantic role; it does not repair missing identity, wrong tool choice,
omitted arguments, or structurally different call representations.

\begin{table}[t]
\centering
\caption{Scope of canonicalization in \textsc{bfcl-canonical}.}
\label{tab:canon_scope}
\scriptsize\setlength{\tabcolsep}{2pt}
\begin{tabular}{p{3.5cm}cp{2.5cm}}
\toprule
Variation & Can.? & Rationale \\
\midrule
Func-name key: \mbox{\texttt{name}} / \mbox{\texttt{function}} / \mbox{\texttt{function\_name}} & Yes & Same semantic role \\
Arg key: \mbox{\texttt{arguments}} / \mbox{\texttt{parameters}} / \mbox{\texttt{params}} & Yes & Same semantic role \\
Func-name key: \mbox{\texttt{function\_call}} & No & Different structure \\
Missing wrapper & No & Identity absent \\
Wrong function name & No & Changes semantics \\
Argument key order & Yes & Order irrelevant \\
Argument omission & No & Changes execution \\
\bottomrule
\end{tabular}
\end{table}

\paragraph{Attribution rule.}
We use the Figure~\ref{fig:attribution_protocol} terminology throughout. We attribute
a gain to \emph{interface alignment} when format-only already accounts for the
main score movement. We call a residual a \emph{procedure-supported residual gain}
only when it remains positive after canonicalization, exceeds format-only,
neutral, and length-matched generic controls, survives the \emph{induction
robustness} checks, and is not explained solely by improved wrapper compliance.
Finally, the \emph{portability} check asks whether that residual keeps its
direction under a new output contract or a cross-benchmark evaluation; only then
do we make a transfer claim. These terms state the evidence required for each
claim strength; they do not identify independent latent mechanisms. We use
$\pm3$~pp as the descriptive neutral band and prefer paired tests or confidence
intervals when available; Appendix~\ref{sec:appendix_threshold} shows the
harmful-cell diagnosis is stable under $\pm2$, $\pm3$, and $\pm5$~pp bands.
Reported cells are single-run unless marked as paired or extraction-seed
analyses, and Appendix~\ref{sec:appendix_audit} summarizes their source files.

For API-Bank, the primary local metric checks the predicted API name and
required ground-truth parameters under API-Bank's bracket syntax. Because it
does not penalize all extra parameters, we call it native API/required-parameter
match rather than official leaderboard accuracy; main API-Bank claims must
also agree qualitatively under strict exact-parameter and row-fresh diagnostics.
All API-Bank target-native comparisons use a source-file-held-out split with
the same examples across skill, format, content, and generic controls.

Appendix Table~\ref{tab:fullmatrix} shows the complete 8$\times$5 BFCL
accuracy matrix under standard scoring.

\section{Structured-Output Gain Attribution in Function Calling}
\label{sec:results}

We organize the results around four research questions (RQs). \textbf{RQ1}
asks how much observed BFCL gain comes from scorer/interface sensitivity rather
than better call selection. \textbf{RQ2} asks whether direct interface contracts
can match or exceed full extracted skills. \textbf{RQ3} asks whether the
largest gains survive repaired and balanced induction. \textbf{RQ4} asks
whether the diagnosis persists across model families and output contracts.
Three findings support the main claim: direct format contracts can match or
exceed full skills; repaired/balanced induction removes the largest gains; and
API-Bank generic procedural prompts nearly reproduce extracted-skill gains.
Canonicalized rescoring, cross-family checks, and selected BFCL official-prompt
non-live controls set the scope of the claim.
Table~\ref{tab:attribution-ledger} summarizes these evaluation settings and the
claims they support before the detailed diagnostics.

\begin{table*}[t]
\centering
\caption{Main evidence map and allowed claims. Detailed numeric estimates and paired tests are in Tables~\ref{tab:scoring}--\ref{tab:repaired} and the appendix.}
\label{tab:attribution-ledger}
\small
\setlength{\tabcolsep}{3pt}
\renewcommand{\arraystretch}{1.12}
\begin{tabular}{@{}>{\raggedright\arraybackslash}p{2.5cm}
>{\raggedright\arraybackslash}p{3.5cm}
>{\raggedright\arraybackslash}p{5.2cm}
>{\raggedright\arraybackslash}p{3.4cm}@{}}
\toprule
Evidence slice & What it tests & Verdict for attribution & Boundary \\
\midrule
BFCL scorer sensitivity & Standard vs.\ canonicalized all-call AST scoring on the same outputs. & Some apparent gains or harms are scorer-sensitive rather than task-policy changes. & Companion diagnostic, not an official leaderboard metric. \\
BFCL format control & Full extracted skill vs.\ concise format-only prompts. & Format-only can match or exceed full skill, so the full-skill gain is not uniquely procedural. & Does not imply format compliance is unimportant. \\
BFCL induction control & Original high-score induction vs.\ balanced/repaired induction. & Flagship sub-frontier gains do not survive cleaner induction controls. & Diagnoses extraction/demo sensitivity, not a single latent cause. \\
Official-prompt BFCL checks & Upstream prompt/evaluator on selected non-live BFCL suites. & No reliable extracted-skill advantage for the main Qwen/Llama checks; Gemma remains a boundary residual. & Selected non-live audits, not full BFCL leaderboard runs. \\
API-Bank target-native checks & Held-out API-Bank rows with native syntax and matched generic prompts. & Generic procedural prompts nearly reproduce target-native skill gains. & Local diagnostics, not official API-Bank leaderboard accuracy. \\
\bottomrule
\end{tabular}
\end{table*}

\subsection{Diagnostic: Evaluator Sensitivity to JSON Key Aliases}
\label{sec:format_aliasing}

Before attributing gains to procedural transfer, we test evaluator robustness.
BFCL-style scoring can distinguish two outputs that choose the same function
and arguments but use different wrapper keys, so key choice can change the
standard score without changing the call decision.

\begin{figure}[t]
\centering
\includegraphics[width=\columnwidth]{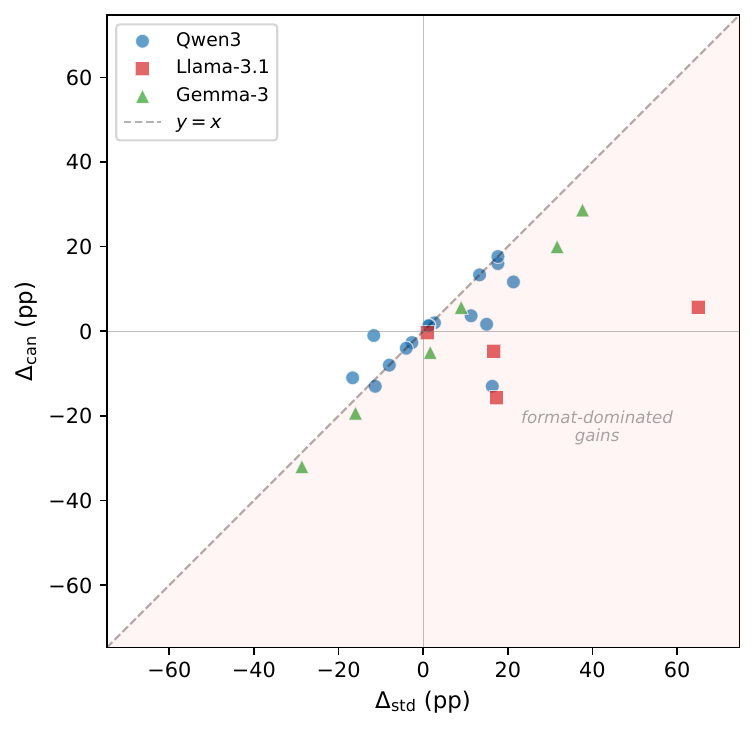}
\caption{Standard-scoring gain vs.\ canonicalized gain for all BFCL model--condition pairs. Points far from the diagonal identify cells where wrapper-key sensitivity changes the interpretation of a skill gain.}
\label{fig:scatter}
\end{figure}

\paragraph{Canonicalized rescoring.}
In our BFCL-v2 JSON scoring setup, the standard scorer treats
\texttt{"name"} as the function-identifier key and marks
\texttt{"function"} or \texttt{"function\_name"} wrong even when arguments are
correct. We define \emph{canonicalized scoring} as AST accuracy after
normalizing key variants (Eq.~\ref{eq:canonical}) and decompose
standard-scoring gain as:
\begin{equation}
  \label{eq:decomp}
  \underbrace{\Delta_{\mathrm{std}}}_{\text{observed gain}}
  = \underbrace{(\Delta_{\mathrm{std}} - \Delta_{\mathrm{can}})}_{\text{alias-sensitive component}}
  + \underbrace{\Delta_{\mathrm{can}}}_{\text{alias-robust component}}
\end{equation}
where deltas are skill-minus-\none{} changes. This estimates key-alias
sensitivity, not all formatting effects (Figure~\ref{fig:scatter}). Using
$\pm3$~pp as the harmful/neutral threshold, 3 of 15 harmful cells in the
full $8 \times 4$ matrix flip to neutral under canonicalization
(Table~\ref{tab:full_std_can}). Canonicalization removes the apparent harm
for 14B \skill{235B} ($-11.7 \to -1.0$~pp) and shrinks the 8B
\skill{235B} gain ($+11.3 \to +3.7$~pp), but residual harms remain: 4B
\skill{235B} stays negative ($-16.7 \to -11.0$~pp) and 32B
\textsc{BFCL-Contract} remains $-17.0$~pp. Across 8 models, format-shift
magnitude predicts standard-scoring gain:
\begin{equation}
  \label{eq:regression}
  \Delta_{\mathrm{std}} \approx \beta_0 + \beta_1 \cdot \delta_m, \quad r = 0.84,\; R^2 = 0.71
\end{equation}
(Pearson $r=0.84$, $n=8$, $p=0.009$; descriptive given sample size),
accounting for 71\% of \textsc{BFCL-Contract} gain variance. Together,
these diagnostics indicate real key-alias sensitivity, but canonicalization is
only a companion diagnostic, not a complete semantic evaluator or the main
mechanism claim. The next tests therefore ask whether full skills add value
beyond the interface contract itself.

\begin{table}[t]
\centering
\caption{BFCL standard vs.\ canonicalized scoring ($\Delta$ pp vs.\ None). Fmt = accuracy restricted to outputs using \texttt{"name"}; bold cells are harmful standard-score deltas that become neutral after canonicalization.}
\label{tab:scoring}
\scriptsize
\setlength{\tabcolsep}{2.5pt}
\begin{tabular*}{\columnwidth}{@{\extracolsep{\fill}}lrrrr@{}}
\toprule
\multicolumn{5}{c}{BFCL-contract prompt} \\
\midrule
Target & None & Std & Can & Fmt \\
\midrule
0.6B      & 26.0 & \ensuremath{+19.3} & \ensuremath{+17.7} & \ensuremath{-7.3} \\
1.7B      & 50.7 & \ensuremath{+13.3} & \ensuremath{+12.7} & \ensuremath{+1.6} \\
4B        & 42.0 & \ensuremath{+20.3} & \ensuremath{-9.0} & \ensuremath{-14.5} \\
8B        & 58.0 & \ensuremath{+14.3} & \ensuremath{+4.7} & \ensuremath{-18.5} \\
14B       & 64.3 & \ensuremath{+3.0} & \ensuremath{-10.3} & \ensuremath{-24.3} \\
32B       & 77.7 & \ensuremath{-17.0} & \ensuremath{-17.0} & \ensuremath{-17.6} \\
30B-A3B   & 70.0 & \ensuremath{-1.7} & \ensuremath{-1.7} & \ensuremath{-1.7} \\
235B      & 80.7 & \ensuremath{-7.7} & \ensuremath{-7.7} & \ensuremath{-9.4} \\
\midrule
\multicolumn{5}{c}{\skill{235B}} \\
\midrule
Target & None & Std & Can & Fmt \\
\midrule
0.6B      & 26.0 & \ensuremath{-2.7} & \ensuremath{-2.7} & \ensuremath{+8.9} \\
1.7B      & 50.7 & \ensuremath{+17.7} & \ensuremath{+17.7} & \ensuremath{+7.4} \\
4B        & 42.0 & \ensuremath{-16.7} & \ensuremath{-11.0} & \ensuremath{-19.6} \\
8B        & 58.0 & \ensuremath{+11.3} & \ensuremath{+3.7} & \ensuremath{-2.7} \\
14B       & 64.3 & \textbf{\ensuremath{-11.7}} & \textbf{\ensuremath{-1.0}} & \ensuremath{-0.4} \\
32B       & 77.7 & \ensuremath{-11.3} & \ensuremath{-13.0} & \ensuremath{-8.0} \\
30B-A3B   & 70.0 & \ensuremath{+1.3} & \ensuremath{+1.3} & \ensuremath{+0.9} \\
235B      & 80.7 & \ensuremath{-4.0} & \ensuremath{-4.0} & \ensuremath{-5.0} \\
\bottomrule
\end{tabular*}
\end{table}

\paragraph{Format-only control.}
We isolate format effects with a concise format-only JSON specification that
states the expected wrapper without task procedure. For Qwen3-8B, format-only
gains $+21.3$~pp, exceeding full \skill{235B} ($+11.3$~pp). For 14B it is also
positive ($+15.0$~pp) while the full skill is harmful; for 32B it is harmful
($-8.0$~pp), showing that interface instructions are sufficient to account for
major score movement in some cells but remain model-state dependent. Alias and
schema controls show the same pattern, so we treat format-only $\geq$ full
skill as evidence that the full-skill gain is not uniquely attributable to
extracted procedural content.

\begin{figure*}[t]
\centering
\includegraphics[width=0.82\textwidth]{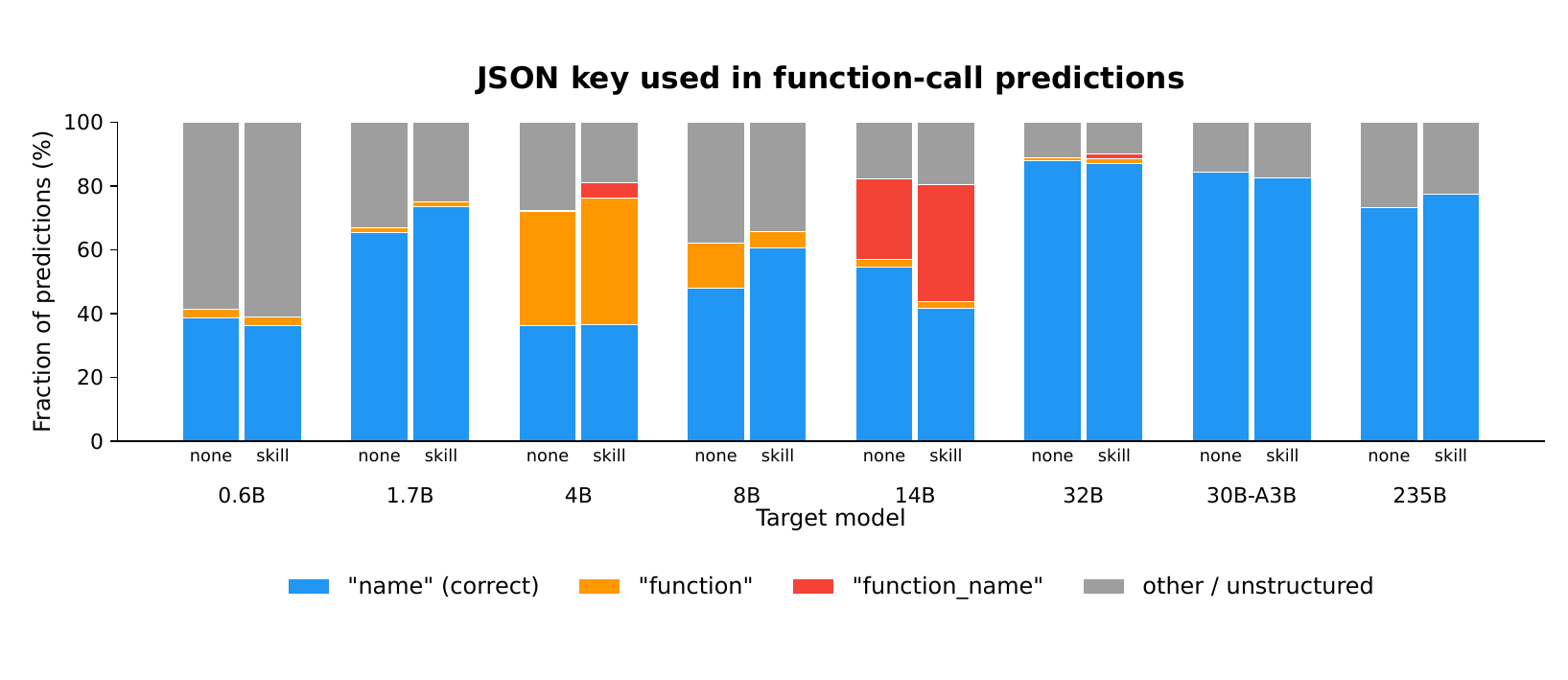}
\caption{BFCL wrapper-key distributions before and after \skill{235B}. For each target model, the no-skill and \skill{235B} bars are shown as an adjacent pair. Blue \texttt{"name"} denotes the expected wrapper key; the figure shows wrapper-key distributions rather than overall accuracy. Skill text can move models toward or away from the expected key, so wrapper compliance must be separated from tool-choice correctness.}
\label{fig:format_dist}
\end{figure*}

\paragraph{Mechanism.}
Figure~\ref{fig:format_dist} illustrates the format-distribution mechanism directly:
for 14B,
\skill{235B} redirects correct calls toward \texttt{"function\_name"}; among
48 standard-score regressions, 35 become correct after canonicalized scoring.

\subsection{Interface Contracts versus Full Skills}
\label{sec:mechanism}
\label{sec:repaired}

We next test whether injected skills exploit evaluator sensitivity by acting
as format instructions rather than procedural knowledge.

\paragraph{Repaired/balanced controls weaken the largest gains.}
Repaired extraction changes extractor family and induction pool, so it is not a
single-factor causal isolation. We use it as an attribution stress test:
Table~\ref{tab:repaired} shows that the largest original positives do not
survive this repaired-and-balanced stress test. Because the extractor, prompt,
and demonstration pool
change together, a reversal supports sensitivity to induction design but does
not identify one causal factor.

For Qwen3-8B, the original $+11.3$~pp gain becomes $-7.7$~pp after repair;
for 1.7B, $+17.7$~pp shrinks to $+3.3$~pp. 235B is slightly negative under
both original and repaired skills ($-4.0$ and $-3.3$~pp), and 4B, 14B, and
32B are negative under both. Appendix~\ref{sec:case_studies} gives concrete
item-level examples where the original skill succeeds through wrapper cues
that the repaired skill no longer supplies.

\begin{table}[t]
\centering
\caption{Effect of \skill{235B} before and after extraction repair on BFCL. Original skills use GPT-4o extraction from the biased induction pool; repaired skills use Qwen3-32B extraction from a balanced pool. Repaired effects are mixed rather than uniformly removed, so they diagnose extraction/demo sensitivity rather than proving a single mechanism.}
\label{tab:repaired}
\footnotesize\setlength{\tabcolsep}{3pt}
\begin{tabular}{l r r r r}
\toprule
Target & None & Original & Repaired & $\Delta$ Change \\
\midrule
0.6B & 26.0 & \ensuremath{-2.7} & \ensuremath{-5.3} & \ensuremath{-2.7} \\
1.7B & 50.7 & \ensuremath{+17.7} & \ensuremath{+3.3} & \ensuremath{-14.3} \\
4B & 42.0 & \ensuremath{-16.7} & \ensuremath{-19.0} & \ensuremath{-2.3} \\
8B & 58.0 & \ensuremath{+11.3} & \ensuremath{-7.7} & \ensuremath{-19.0} \\
14B & 64.3 & \ensuremath{-11.7} & \ensuremath{-9.7} & \ensuremath{+2.0} \\
32B & 77.7 & \ensuremath{-11.3} & \ensuremath{-10.3} & \ensuremath{+1.0} \\
30B-A3B & 70.0 & \ensuremath{+1.3} & \ensuremath{-3.7} & \ensuremath{-5.0} \\
235B & 80.7 & \ensuremath{-4.0} & \ensuremath{-3.3} & \ensuremath{+0.7} \\
\bottomrule
\end{tabular}
\end{table}

\paragraph{Balanced induction.}
The extractor-by-demonstration comparisons vary source/extractor style and
high-score demonstration selection, but do not isolate them as independent
causes. For 8B, biased
pools help under both GPT-4o and Qwen3-32B extraction ($+11.3$ and
$+11.0$~pp), while balanced demos reverse both. For 1.7B, $+17.7$~pp falls to
$-1.7$~pp with balanced GPT-4o subsets and $+3.3$~pp with repaired Qwen3
extraction; 14B is sign-unstable. These results are consistent with
induction-design sensitivity: the strongest positive
gains depend on the particular high-scoring induction examples and extraction
setup, rather than demonstrating a stable extracted procedure.

\paragraph{Residual controls and subtask heterogeneity.}
\label{sec:subtask}
Content-only and subtask controls are consistent with this diagnosis but are
not central to the evidence. Content-only prompting helps weaker models and
hurts 14B, while Table~\ref{tab:category} shows that skill injection helps
\texttt{simple\_python} for 1.7B/8B but harms it for 4B/14B and leaves
parallel-call items brittle. The pattern is useful for localization, but it is
not uniform procedural transfer.

\subsection{Demonstration Selection and Balanced Induction}
\label{sec:attribution}
\label{sec:induction}

The scorer is sensitive to wrapper format, and skills can change wrapper
choice. We now ask why the original extraction pipeline produced useful format
cues.

\paragraph{Induction sampling bias.}
Original GPT-4o skills sampled high-scoring \none{} demonstrations,
over-representing format-compliant calls. We
quantify this with three biased and three balanced subset runs
(Appendix~\ref{sec:appendix_variance}).

\begin{table}[t]
\centering
\caption{Effect of induction sampling on extracted skill quality (BFCL, \skill{235B} source). Original uses the Phase~1 GPT-4o biased pool. Biased and balanced entries are means over available subset runs; repaired uses Qwen3-32B extraction from a balanced pool.}
\label{tab:induction}
\footnotesize\setlength{\tabcolsep}{3pt}
\begin{tabular}{l r r r r}
\toprule
Target & Original & Biased (mean) & Balanced (mean) & Repaired \\
\midrule
1.7B & \ensuremath{+17.7} & \ensuremath{-6.3} & \ensuremath{-1.7} & \ensuremath{+3.3} \\
8B & \ensuremath{+11.3} & \ensuremath{+3.2} & \ensuremath{-3.8} & \ensuremath{-7.7} \\
32B & \ensuremath{-11.3} & \ensuremath{-6.2} & \ensuremath{-6.6} & \ensuremath{-10.3} \\
235B & \ensuremath{-4.0} & \ensuremath{-6.7} & \ensuremath{-5.3} & \ensuremath{-3.3} \\
\bottomrule
\end{tabular}
\end{table}

For 8B, biased subsets average $+3.2$~pp while balanced subsets average
$-3.8$~pp. For 1.7B, five of six subsets are negative and the sixth is only
$+4.0$~pp, far below the original $+17.7$~pp. Qwen3-235B also lacks a
positive balanced-induction residual (mean $-5.3$~pp). The large original
gains are therefore more consistent with sensitivity to the induction design than with a stable
extracted procedure. The subset resampling results complement, but do not by
themselves identify, the source of that sensitivity.

\subsection{Boundary Conditions and Cross-Contract Checks}
\label{sec:boundary}
\label{sec:crossfamily}

\paragraph{Cross-family and official-prompt BFCL.}
On non-Qwen BFCL, format-only exceeds the Qwen3-derived full skill in all five
rows (Table~\ref{tab:cross_family}); Llama-3.1-70B gains $+65.0$~pp from the
format cue but only $+1.0$~pp from \skill{235B}, while the Qwen3-derived skill
reduces scores for smaller Gemma models. The official-prompt non-live check in
Table~\ref{tab:attribution-ledger} preserves the boundary interpretation:
Qwen3-8B and Llama-3.1-8B show no reliable extracted-skill advantage, whereas
Gemma-3-4B retains a model-specific residual. Appendix
Section~\ref{sec:appendix_official_controls} adds content-only Qwen,
prompt-variant, repaired-source, and same-family Llama checks; none yields a
stable residual beyond matched controls.

\paragraph{Second benchmark: API-Bank.}
\label{sec:apibank}
API-Bank~\citep{li2023apibank} uses \texttt{[ApiName(params)]}. Native format
cues have almost no mean effect ($+0.3$~pp), whereas the BFCL JSON cue reduces
native API/required-parameter match by 19.4--54.8~pp
(Appendix~\ref{sec:appendix_apibank}). Under the target-native contract, the extracted
skill improves over native format-only, but length-matched generic procedural
prompts and generic variants nearly match it under held-out source files. Thus,
API-Bank provides a local cross-contract diagnostic rather than an official
leaderboard result.

\paragraph{Scope checks: reasoning and QA.}
\label{sec:null}
\label{sec:brevity}
MATH-500 and MultiHop-RAG are mostly null scope checks: MATH-500 stays within
$\pm 3$~pp for all 8 models, and MultiHop-RAG stays within $\pm 4$~pp for 7/8
models, with one Qwen3-235B exception.

\section{Discussion}
\label{sec:discussion}

\paragraph{Unified account and scope.}
BFCL key-alias sensitivity creates room for format-driven score changes; skills
can act as implicit format instructions; and high-scoring demonstration
selection amplifies those cues. Direct interface prompts, repaired/balanced
induction, and generic API-Bank prompts help
account for much of the measured movement. Together, these are stress
tests of the attribution claim, not interventions on independent latent
modules; the format-only result is a sufficiency check rather than evidence
that procedural behavior is absent. The protocol prevents unsupported transfer
claims and specifies the additional evidence required before a residual can be
credited to procedure.
The official-prompt Gemma residual remains a boundary case. Appendix
Table~\ref{tab:spider_attribution_m23} reports an auxiliary Spider
check with the same interface-compliance pattern.

\paragraph{What the claim is.}
A skill-injection gain should not be credited to procedural transfer merely
because a skill beats \none{} under one scorer. The stronger standard is whether
the effect survives canonicalization, format/procedural controls, balanced or
repaired induction, and target-native or cross-family checks. This does not
discount format compliance, which is a useful engineering capability; it says that
gains failing these checks should be reported as interface alignment or
prompt-contract adaptation rather than transferable procedure.

\section{Conclusion}
\label{sec:conclusion}

Structured-output gain attribution clarifies what a function-calling score can
certify. In our audits, several large prompt-induced gains are real
under the scored interface but do not survive the checks needed for procedural
transfer: format-only controls reproduce key BFCL gains, repaired and balanced
induction remove the largest sub-frontier improvements, and API-Bank generic
procedural prompts nearly match target-native extracted skills. We recommend
reporting canonicalized metrics, format-only baselines, induction-robustness
checks, and portability checks before calling a structured-output gain
transferable skill. Interface compliance is part of building reliable
tool-using systems, and a prompt that teaches the expected wrapper can be
valuable in deployment. The attribution error is to interpret that gain as
evidence that an extracted procedure transferred across models, contracts, or
benchmarks. Our protocol preserves the engineering value of format alignment
while reserving the stronger procedural claim for residual evidence that
survives the controls. These results do not invalidate BFCL or API-Bank;
they show that gains under a fixed contract require attribution before being
interpreted as transferable procedure.

\section*{Acknowledgments}

This work was supported by Project 62276178 under the National Natural Science
Foundation of China, Key Project 23KJA520012 under the Natural Science
Foundation of Jiangsu Higher Education Institutions, Project 24BTQ002 under the
National Social Science Foundation of China, Project 22YJCZH091 of the
Humanities and Social Science Fund of the Ministry of Education, and the
Priority Academic Program Development of Jiangsu Higher Education Institutions.

\section*{Limitations}

We study prompt-prepended text skills, wrapper-key aliasing, two-Qwen schema
rotation, one non-BFCL API-Bank contract, Qwen3-8B/14B instruction-matched
API-Bank baselines, and within-task MATH-500/MultiHop-RAG checks. API-Bank
official-style checks remain local diagnostics because unsupported-execution
rows are excluded; cross-family tests cover Qwen3, Llama-3.1, and Gemma-3; the
full extractor-by-demo design covers 1.7B/8B/14B; and cells are single-run
unless marked as extraction-seed or prompt-mode paired analyses. We therefore
claim an attribution protocol and a scoped BFCL/API-Bank diagnosis, not a
universal result about all structured-output benchmarks. The Spider check in
Appendix~\ref{sec:appendix_spider} is included only as an external
structured-output check, not as a text-to-SQL leaderboard result or a
fourth main benchmark claim.

\bibliography{references}

\appendix
\section{Full Experimental Results}
\label{sec:appendix_full}

This appendix records provenance and stress tests rather than a second results
narrative. Sections~\ref{sec:appendix_fullmatrix_std}--\ref{sec:appendix_crossfamily}
support the BFCL attribution chain; Sections~\ref{sec:appendix_apibank}--\ref{sec:appendix_official_controls}
bound API-Bank and official-prompt scope; the remaining sections document
variance, error types, compute budget, and provenance for prose summaries. The main claim does not
require treating every appendix table as an independent positive result.

\paragraph{Appendix roadmap.}
\begin{description}[style=unboxed,leftmargin=0pt,itemsep=1pt]
\item[\S\ref{sec:appendix_fullmatrix_std}--\S\ref{sec:appendix_category}]
BFCL standard matrix, format-leakage ablations, canonicalized matrices,
scorer-mechanism figures, and per-category effects.
\item[\S\ref{sec:appendix_scope_probes}--\S\ref{sec:appendix_crossfamily}]
MATH/MultiHop scope checks, Spider external structured-output check,
transitions, induction variance, skill examples, and cross-family checks.
\item[\S\ref{sec:appendix_apibank}--\S\ref{sec:appendix_apibank_native}]
API-Bank target-native results, generic-control variants, strict metrics,
fresh-execution diagnostics, unsupported-row extensions, and non-Qwen boundary
checks.
\item[\S\ref{sec:appendix_position}--\S\ref{sec:appendix_official_controls}]
Prompt-position, threshold, error-type decomposition, and BFCL official-prompt
controls.
\item[\S\ref{sec:appendix_seed_variance}--\S\ref{sec:appendix_audit}]
Extraction-seed variance, model-size/compute reporting, reproducibility
checklist, and provenance for prose claims.
\item[Appendix~\ref{sec:case_studies}]
Item-level examples for key-alias sensitivity, format-only sufficiency, and
original-versus-repaired skill behavior.
\end{description}

\subsection{Full BFCL Standard Matrix}
\label{sec:appendix_fullmatrix_std}

Table~\ref{tab:fullmatrix} reports the complete BFCL standard all-call scoring
matrix that underlies the main-text attribution diagnostics.

\begin{table*}[t]
\centering
\caption{Full BFCL accuracy matrix on our fixed 300-item BFCL-v2 JSON subset (standard all-call scoring, \%, 8 models $\times$ 5 conditions). Skill@\{8B,32B,235B\} inject trajectory-derived skills from the corresponding source model. ``Contract'' is our BFCL-style JSON prompt specifying the \texttt{name}/\texttt{arguments} wrapper.}
\label{tab:fullmatrix}
\footnotesize\setlength{\tabcolsep}{3pt}
\begin{tabular}{l r r r r r}
\toprule
Target & None & Contract & skill@8B & skill@32B & skill@235B \\
\midrule
0.6B & 26.0 & 45.3 & 13.3 & 13.3 & 23.3 \\
1.7B & 50.7 & 64.0 & 69.3 & 66.3 & 68.3 \\
4B & 42.0 & 62.3 & 21.3 & 22.7 & 25.3 \\
8B & 58.0 & 72.3 & 67.3 & 68.3 & 69.3 \\
14B & 64.3 & 67.3 & 54.3 & 56.3 & 52.7 \\
32B & 77.7 & 60.7 & 69.3 & 64.3 & 66.3 \\
30B-A3B & 70.0 & 68.3 & 72.3 & 71.3 & 71.3 \\
235B & 80.7 & 73.0 & 77.0 & 78.0 & 76.7 \\
\bottomrule
\end{tabular}
\end{table*}

Table~\ref{tab:bfcl_local_paired_claims} adds paired tests and confidence
intervals for the local BFCL comparisons that support the main attribution
claims: apparent skill gains, format-only versus full skill, and original
versus repaired extraction.

\begin{table*}[t]
\centering
\caption{Paired statistics for the claim-bearing local BFCL comparisons. Each row compares condition A with condition B on the same 300 BFCL rows under standard all-call scoring. A-only and B-only count discordant examples; $p$ is a two-sided exact sign test and CI is a paired bootstrap interval over item-level differences.}
\label{tab:bfcl_local_paired_claims}
\scriptsize\setlength{\tabcolsep}{3pt}
\begin{tabular}{@{}lrrrrrrr@{}}
\toprule
Comparison (A vs B) & A acc. & B acc. & $\Delta$ pp & 95\% CI & A-only & B-only & $p$ \\
\midrule
1.7B skill vs none & 68.3 & 50.7 & $+17.7$ & $[+12.7,+23.0]$ & 62 & 9 & $<.001$ \\
1.7B repaired vs skill & 54.0 & 68.3 & $-14.3$ & $[-19.3,-9.7]$ & 8 & 51 & $<.001$ \\
8B skill vs none & 69.3 & 58.0 & $+11.3$ & $[+7.3,+15.7]$ & 39 & 5 & $<.001$ \\
8B format-only vs skill & 79.3 & 69.3 & $+10.0$ & $[+5.7,+14.3]$ & 39 & 9 & $<.001$ \\
8B repaired vs skill & 50.3 & 69.3 & $-19.0$ & $[-24.0,-14.3]$ & 5 & 62 & $<.001$ \\
14B format-only vs skill & 79.3 & 52.7 & $+26.7$ & $[+20.3,+33.3]$ & 99 & 19 & $<.001$ \\
14B repaired vs skill & 54.7 & 52.7 & $+2.0$ & $[-3.0,+7.0]$ & 34 & 28 & $0.526$ \\
\bottomrule
\end{tabular}
\end{table*}

\subsection{Format-Leakage Ablation Table}
\label{sec:appendix_ablation}

Table~\ref{tab:ablation} presents the ablation across nine skill conditions
varying in format-leakage level (14B baselines are in Table~\ref{tab:scoring}).
For 8B, the broad gradient demonstrates that format leakage drives standard-scoring gains.
Table~\ref{tab:factorial} decomposes the repaired-extraction reversal into its two components.

\begin{table*}[t]
\centering
\caption{Ablation: BFCL accuracy ($\Delta$ pp vs.\ None) for conditions varying the amount and type of format leakage. The pattern is not monotone: short alias or format prompts can exceed the full skill, while content-only and abstract prompts can still move weaker models. Thus format cues are sufficient for many gains, but residual prompt framing is not zero.}
\label{tab:ablation}
\small\setlength{\tabcolsep}{2.5pt}
\begin{tabular}{l r r r r r r r r r r}
\toprule
Target & None & repaired & content-only & abstract & skill@235B & alias-only & fmt-only & wrong-alias & shuffled & raw\_traj \\
\midrule
1.7B & 50.7 & \ensuremath{+3.3} & \ensuremath{+16.7} & \ensuremath{+15.0} & \textbf{\ensuremath{+17.7}} & \ensuremath{-15.0} & \ensuremath{+2.7} & \ensuremath{-41.0} & \ensuremath{-31.3} & \ensuremath{-4.3} \\
\midrule
8B & 58.0 & \ensuremath{-7.7} & \ensuremath{+10.3} & \ensuremath{+4.7} & \ensuremath{+11.3} & \ensuremath{+15.0} & \textbf{\ensuremath{+21.3}} & \ensuremath{-39.7} & \ensuremath{+12.7} & \ensuremath{+11.7} \\
\midrule
14B & 64.3 & \ensuremath{-9.7} & \ensuremath{-10.7} & -- & \ensuremath{-11.7} & \ensuremath{+12.7} & \textbf{\ensuremath{+15.0}} & \ensuremath{-48.3} & -- & -- \\
\midrule
32B & 77.7 & \ensuremath{-10.3} & -- & \textbf{\ensuremath{-7.7}} & \ensuremath{-11.3} & -- & \ensuremath{-8.0} & -- & \ensuremath{-12.0} & \ensuremath{-20.3} \\
\midrule
235B & 80.7 & \ensuremath{-3.3} & -- & \ensuremath{-7.7} & \ensuremath{-4.0} & -- & \textbf{\ensuremath{+1.3}} & -- & \ensuremath{-6.7} & \ensuremath{-12.0} \\
\bottomrule
\end{tabular}
\end{table*}

\begin{table}[t]
\centering
\caption{2$\times$2 factorial decomposition (BFCL, $\Delta$~pp vs.\ none). GPT-4o balanced entries are means over three balanced subsets; Qwen3-32B balanced entries are repaired extractions from a balanced pool. Demo selection has model-dependent effects after all-call rescoring.}
\label{tab:factorial}
\footnotesize\setlength{\tabcolsep}{3pt}
\begin{tabular}{llcc}
\toprule
 & & \textbf{Biased} & \textbf{Balanced} \\
\midrule
\multirow{2}{*}{\textbf{1.7B}} & GPT-4o & $+17.7$ & $-1.7$ \\
 & Qwen3-32B & $+13.7$ & $+3.3$ \\
\addlinespace
\multirow{2}{*}{\textbf{8B}} & GPT-4o & $+11.3$ & $-3.8$ \\
 & Qwen3-32B & $+11.0$ & $-7.7$ \\
\addlinespace
\multirow{2}{*}{\textbf{14B}} & GPT-4o & $-11.7$ & $-3.7$ \\
 & Qwen3-32B & $+4.7$ & $-9.7$ \\
\midrule
\multicolumn{2}{l}{\textit{Mean demo effect (1.7B)}} & \multicolumn{2}{c}{$-14.8$\,pp} \\
\multicolumn{2}{l}{\textit{Mean demo effect (8B)}} & \multicolumn{2}{c}{$-16.9$\,pp} \\
\multicolumn{2}{l}{\textit{Mean demo effect (14B)}} & \multicolumn{2}{c}{$-3.2$\,pp} \\
\bottomrule
\end{tabular}
\end{table}

\subsection{Full Standard vs.\ Canonicalized Matrix}
\label{sec:appendix_full_matrix}

Table~\ref{tab:full_std_can} reports standard and canonicalized $\Delta$
for all $8 \times 4$ non-none BFCL conditions. Bold cells are harmful under
standard scoring ($\Delta_{\mathrm{std}} \le -3$~pp) but flip to neutral
($|\Delta_{\mathrm{can}}| < 3$~pp) under canonicalization.

\begin{table*}[t]
\centering
\caption{Full BFCL standard vs.\ canonicalized scoring matrix ($\Delta$ pp vs.\ None, 8 models $\times$ 4 conditions). \textbf{Bold} = harmful standard-score cells ($\Delta_{\mathrm{std}}\le -3$~pp) that flip to neutral ($|\Delta_{\mathrm{can}}|<3$~pp) under canonicalization. Total: 15 harmful cells, 3 flip (20\%).}
\label{tab:full_std_can}
\scriptsize\setlength{\tabcolsep}{2pt}
\begin{tabular}{l rr rr rr rr}
\toprule
 & \multicolumn{2}{c}{Contract} & \multicolumn{2}{c}{sk@8B} & \multicolumn{2}{c}{sk@32B} & \multicolumn{2}{c}{sk@235B} \\
\cmidrule(lr){2-3}\cmidrule(lr){4-5}\cmidrule(lr){6-7}\cmidrule(lr){8-9}
Target & Std & Can & Std & Can & Std & Can & Std & Can \\
\midrule
0.6B & \ensuremath{+19.3} & \ensuremath{+17.7} & \ensuremath{-12.7} & \ensuremath{-13.3} & \ensuremath{-12.7} & \ensuremath{-13.3} & \ensuremath{-2.7} & \ensuremath{-2.7} \\
1.7B & \ensuremath{+13.3} & \ensuremath{+12.7} & \ensuremath{+18.7} & \ensuremath{+18.7} & \ensuremath{+15.7} & \ensuremath{+15.7} & \ensuremath{+17.7} & \ensuremath{+17.7} \\
4B & \ensuremath{+20.3} & \ensuremath{-9.0} & \ensuremath{-20.7} & \ensuremath{-10.3} & \ensuremath{-19.3} & \ensuremath{-12.0} & \ensuremath{-16.7} & \ensuremath{-11.0} \\
8B & \ensuremath{+14.3} & \ensuremath{+4.7} & \ensuremath{+9.3} & \ensuremath{+2.7} & \ensuremath{+10.3} & \ensuremath{+3.7} & \ensuremath{+11.3} & \ensuremath{+3.7} \\
14B & \ensuremath{+3.0} & \ensuremath{-10.3} & \textbf{\ensuremath{-10.0}} & \textbf{\ensuremath{+0.7}} & \textbf{\ensuremath{-8.0}} & \textbf{\ensuremath{-0.7}} & \textbf{\ensuremath{-11.7}} & \textbf{\ensuremath{-1.0}} \\
32B & \ensuremath{-17.0} & \ensuremath{-17.0} & \ensuremath{-8.3} & \ensuremath{-8.3} & \ensuremath{-13.3} & \ensuremath{-14.0} & \ensuremath{-11.3} & \ensuremath{-13.0} \\
30B-A3B & \ensuremath{-1.7} & \ensuremath{-1.7} & \ensuremath{+2.3} & \ensuremath{+2.3} & \ensuremath{+1.3} & \ensuremath{+1.3} & \ensuremath{+1.3} & \ensuremath{+1.3} \\
235B & \ensuremath{-7.7} & \ensuremath{-7.7} & \ensuremath{-3.7} & \ensuremath{-3.7} & \ensuremath{-2.7} & \ensuremath{-2.7} & \ensuremath{-4.0} & \ensuremath{-4.0} \\
\bottomrule
\end{tabular}
\end{table*}

\subsection{Scorer-Mechanism Figure Notes}
\label{sec:appendix_scorer_figures}

Figures~\ref{fig:scatter} and~\ref{fig:format_dist} are placed in the main text
because they are the two most direct visual diagnostics for scorer sensitivity:
the scatter gives the global standard-vs-canonicalized view, and the
key-distribution figure shows the wrapper-key mechanism.

\subsection{Per-Category BFCL Effects}
\label{sec:appendix_category}

Table~\ref{tab:category} decomposes the \skill{235B} effect by BFCL category.
It supports the main-text claim that effects are subtask-selective rather than
uniform procedural transfer.

\begin{table}[t]
\centering
\caption{Per-category effect of skill@235B on BFCL under standard all-call scoring. Effects are subtask-selective rather than uniform, consistent with format and prompt steering.}
\label{tab:category}
\small
\begin{tabular}{l l r r r}
\toprule
Target & Category & None & skill@235B & $\Delta$ \\
\midrule
1.7B & simple\_python & 56.6 & 88.7 & \ensuremath{+32.1} \\
 & multiple & 85.5 & 79.0 & \ensuremath{-6.5} \\
 & parallel & 8.1 & 14.5 & \ensuremath{+6.5} \\
 & irrelevance & 48.6 & 75.7 & \ensuremath{+27.1} \\
\midrule
8B & simple\_python & 63.2 & 89.6 & \ensuremath{+26.4} \\
 & multiple & 79.0 & 85.5 & \ensuremath{+6.5} \\
 & parallel & 8.1 & 0.0 & \ensuremath{-8.1} \\
 & irrelevance & 75.7 & 85.7 & \ensuremath{+10.0} \\
\midrule
4B & simple\_python & 25.5 & 12.3 & \ensuremath{-13.2} \\
 & multiple & 85.5 & 77.4 & \ensuremath{-8.1} \\
 & parallel & 9.7 & 4.8 & \ensuremath{-4.8} \\
 & irrelevance & 57.1 & 17.1 & \ensuremath{-40.0} \\
\midrule
14B & simple\_python & 66.0 & 45.3 & \ensuremath{-20.8} \\
 & multiple & 87.1 & 69.4 & \ensuremath{-17.7} \\
 & parallel & 37.1 & 33.9 & \ensuremath{-3.2} \\
 & irrelevance & 65.7 & 65.7 & \ensuremath{+0.0} \\
\bottomrule
\end{tabular}
\end{table}

\subsection{Full MATH-500 and MultiHop-RAG Matrices}
\label{sec:appendix_scope_probes}

Tables~\ref{tab:math_full} and~\ref{tab:mhrag_full} present the full accuracy
matrices for MATH-500 (sympy equivalence accuracy) and MultiHop-RAG (F1),
respectively, for all 8 models $\times$ 5 conditions.
All MATH-500 effects are within $\pm 3$~pp; MultiHop-RAG effects are within
$\pm 4$~pp for 7 of 8 models (the exception is Qwen3-235B
at $+5.5$ to $+7.2$~pp).

\begin{table*}[h]
\centering
\caption{Full MATH-500 accuracy (\%, sympy equivalence, 8 models $\times$ 5 conditions). All effects are within $\pm 3$~pp, consistent with a null effect for symbolic reasoning.}
\label{tab:math_full}
\small
\begin{tabular}{l r r r r r}
\toprule
Target & None & Contract & skill@8B & skill@32B & skill@235B \\
\midrule
0.6B & 41.5 & 40.0 & 41.9 & 43.4 & 41.9 \\
1.7B & 64.3 & 64.5 & 64.5 & 65.5 & 63.4 \\
4B & 73.2 & 72.8 & 73.0 & 72.8 & 73.0 \\
8B & 75.1 & 74.0 & 73.2 & 74.7 & 74.7 \\
14B & 77.7 & 76.4 & 77.9 & 78.3 & 76.4 \\
32B & 78.5 & 77.0 & 79.8 & 77.9 & 78.7 \\
30B-A3B & 78.5 & 75.7 & 79.6 & 79.6 & 79.1 \\
235B & 75.3 & 72.6 & 78.1 & 75.3 & 75.7 \\
\bottomrule
\end{tabular}
\end{table*}

\begin{table*}[h]
\centering
\caption{Full MultiHop-RAG F1 (\%, 8 models $\times$ 5 conditions). Under skill@\{8B,32B,235B\} conditions, 7 of 8 models are within $\pm 4$~pp; the sole exception is Qwen3-235B. The BFCL-contract prompt condition shows larger effects for some models.}
\label{tab:mhrag_full}
\small
\begin{tabular}{l r r r r r}
\toprule
Target & None & Contract & skill@8B & skill@32B & skill@235B \\
\midrule
0.6B & 0.4 & 0.5 & 0.3 & 0.4 & 0.4 \\
1.7B & 15.1 & 1.4 & 15.4 & 14.3 & 12.7 \\
4B & 11.5 & 30.7 & 12.3 & 9.5 & 9.4 \\
8B & 31.6 & 31.3 & 29.7 & 28.4 & 30.0 \\
14B & 31.9 & 31.7 & 31.8 & 31.6 & 32.2 \\
32B & 33.7 & 32.2 & 33.7 & 34.2 & 35.8 \\
30B-A3B & 34.4 & 32.8 & 31.8 & 32.0 & 32.9 \\
235B & 34.4 & 39.3 & 41.6 & 40.4 & 39.9 \\
\bottomrule
\end{tabular}
\end{table*}

\begin{figure}[t]
\centering
\includegraphics[width=\columnwidth]{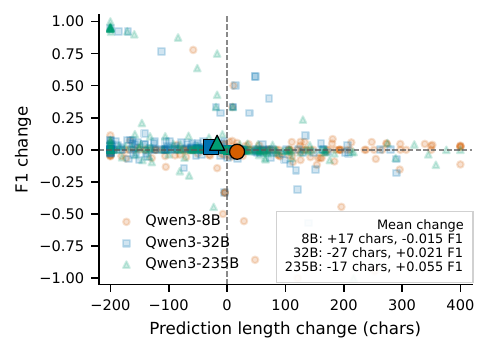}
\caption{Per-item MultiHop-RAG changes under \skill{235B} versus \none{}.
Small points are item-level prediction-length and F1 changes; large markers are
model means. Qwen3-235B is the only notable MultiHop-RAG gain in this scope
check.}
\label{fig:multihop_brevity}
\end{figure}

\subsection{Spider Structured-Output Probe}
\label{sec:appendix_spider}

Table~\ref{tab:spider_attribution_m23} reports a local Spider dev-subset check
that tests whether the attribution protocol's interface-alignment diagnosis
appears beyond function-calling wrappers. The check uses 300 deterministic
Spider rows and Qwen3-8B/14B. \emph{Strict exec} credits only raw SQL-only
outputs; \emph{Extracted-SQL exec} first removes prose or markdown wrappers
before executing the SQL against the SQLite database. This is an external
structured-output check, not an official Spider result.

\begin{table*}[t]
\centering
\small
\begin{tabular}{llrrrr}
\toprule
Model & Condition & $n$ & Strict exec & Extracted-SQL exec & SQL-only output \\
\midrule
14B & Full SQL skill & 300 & 71.7 & 71.7 & 100.0 \\
14B & Extracted skill & 300 & 2.0 & 70.0 & 3.0 \\
14B & Format-only & 300 & 70.7 & 70.7 & 100.0 \\
14B & None & 300 & 44.0 & 70.3 & 58.7 \\
14B & Procedure-only & 300 & 0.0 & 69.7 & 0.0 \\
8B & Full SQL skill & 300 & 67.0 & 67.0 & 100.0 \\
8B & Extracted skill & 300 & 25.7 & 63.0 & 32.0 \\
8B & Format-only & 300 & 64.7 & 64.7 & 100.0 \\
8B & None & 300 & 5.0 & 66.3 & 5.7 \\
8B & Procedure-only & 300 & 2.0 & 66.3 & 2.7 \\
\bottomrule
\end{tabular}
\caption{Spider structured-output attribution probe. Strict execution credits only raw SQL-only outputs; extracted-SQL execution first removes markdown or prose wrappers. The probe is an external structured-output sanity check, not an official Spider leaderboard result.}
\label{tab:spider_attribution_m23}
\end{table*}

The separation is large. For Qwen3-8B, \none{} scores 66.3\% under extracted-SQL
execution but only 5.0\% under strict raw-SQL execution; for Qwen3-14B, the
corresponding values are 70.3\% and 44.0\%. Three independently worded
SQL-only format prompts recover strict accuracy to 64.7--67.0\% for 8B and
70.3--70.7\% for 14B, with 100\% SQL-only interface compliance. By contrast,
procedure-only and extracted procedural prompts preserve extracted-SQL accuracy
but usually fail strict interface compliance: procedure-only yields 66.3\% and
69.7\% extracted-SQL execution for 8B/14B, but 2.0\% and 0.0\% strict
execution. Paired contrasts confirm that format-only is not significantly
below full DIN-SQL skill on strict execution (8B: $-2.3$~pp, 95\% CI
$[-5.0,0.0]$; 14B: $-1.0$~pp, 95\% CI $[-3.0,+1.0]$), while procedure-only is
65.0--71.7~pp below full DIN-SQL skill under strict execution. These results
support the scoped claim that interface-compliance attribution is not unique
to BFCL/API-Bank-style function-call wrappers.

\subsection{Per-Item Transition Analysis}
\label{sec:appendix_transitions}

For each model $\times$ condition on \bfcl{}, we tabulate the number of items
that transition from correct to incorrect (regressions) and vice versa
(improvements) between \none{} and the skill condition.
For MATH-500, regression and improvement counts are approximately balanced
across all models and conditions (average imbalance about $4.5$ items out of 470,
and about $3.7$ items for skill-only comparisons),
consistent with random rather than systematic effects.
For \bfcl{}, wrapper-key switching explains a substantial subset of standard-score
regressions, but the residual errors discussed in
Section~\ref{sec:appendix_error_breakdown} show that canonicalization is not a
complete semantic evaluator.

\subsection{Induction Variance and Subset Deltas}
\label{sec:appendix_variance}

To quantify sensitivity to the particular demonstrations used for skill extraction,
we ran 6 independent induction subsets
for \skill{235B} on four target models.
We define \emph{spread} as max$-$min accuracy across the 6 subsets,
and \emph{variance ratio} as spread / (expected spread under IID binomial sampling,
computed as $2\sigma = 2\sqrt{p(1-p)/n}$ for each model's mean accuracy $p$ and
$n=300$ items).
A ratio $>1$ indicates the extraction subsets produce more variance
than random noise; $<1$ indicates the model is more stable than expected.
The resulting spreads are 11.3~pp for 1.7B (2.0$\times$ random-noise scale),
11.0~pp for 8B (1.9$\times$), 6.0~pp for 32B (1.1$\times$), and 3.3~pp for
235B (0.7$\times$). The individual subset deltas explain the interpretation:
1.7B has original $+17.7$~pp but subset deltas of $-7.3,-5.7,-6.0,-5.3,+4.0$,
and $-3.7$~pp; 8B has original $+11.3$~pp but balanced subsets of $-6.7,-3.0$,
and $-1.7$~pp; 14B is sign-unstable across balanced subsets
($+14.3,-17.3,-8.0$~pp); 32B and 235B stay negative across the available
subsets. This result is therefore summarized in prose rather than printed as a
separate table.

\subsection{Skill Text Examples}
\label{sec:appendix_skill_examples}

To illustrate the format-aliasing mechanism qualitatively,
we show representative excerpts from original (GPT-4o) and repaired (Qwen3-32B)
skills for the function-calling task.

\textbf{Format-only skill (concise control):}
\begin{quote}
\small
\textit{``When making a function call, structure your response as a JSON object with
exactly two top-level keys: `name' (the exact function name, string) and `arguments'
(an object containing the required parameters).
Example: \texttt{\{"name": "function\_name", "arguments": \{"param1": v1\}\}}.
Do not add any other keys or wrapper structure.''}
\end{quote}

\textbf{Original skill (GPT-4o):}
\begin{quote}
\small
\textit{``To complete function calls correctly, structure your response as a JSON object
with \texttt{"name"} and \texttt{"arguments"} keys.
The \texttt{"name"} key must match the function name exactly as specified in the tool
schema. Always include all required parameters in \texttt{"arguments"}...''}
\end{quote}

\textbf{Repaired skill (Qwen3-32B):}
\begin{quote}
\small
\textit{``When the user asks for information that requires a tool call, identify the
correct function from the provided tool list and construct the call with the required
parameters. Ensure parameter types match the function schema.
For complex queries involving multiple conditions, decompose into sequential calls...''}
\end{quote}

The original skill explicitly names the JSON keys (\texttt{"name"}, \texttt{"arguments"}),
directly triggering the format-aliasing effect.
The repaired skill focuses on procedural content (how to identify and construct calls)
without naming the keys, which is why it produces smaller and less format-driven effects.

\paragraph{Additional ablation conditions.}
\emph{Alias-only} (14 words): ``If you output a function call, use the key \texttt{"name"} for the function identifier.''
\emph{Wrong-alias} (14 words): same sentence but requesting \texttt{"function"} instead.
\emph{Content-only}: the full extracted skill with all JSON format tokens (key names, braces, examples) removed, retaining only procedural instructions (200 words).
\emph{Abstract}: a summary of the skill's high-level principles without any JSON key names or format examples.
\emph{Shuffled}: the full skill with step ordering randomized but all key tokens preserved.
\emph{Raw trajectory}: verbatim successful function-call outputs from the source model, without any distillation.

\subsection{Cross-Family Validation}
\label{sec:appendix_crossfamily}

Table~\ref{tab:cross_family} presents the full cross-family results discussed
in Section~\ref{sec:crossfamily}, showing that format-only outperforms
the full skill across the seven-model suite under standard scoring,
and that Qwen3-extracted skills reduce scores for smaller Gemma models.

\begin{table*}[t]
\centering
\caption{\textbf{Cross-family validation on BFCL.} Qwen3-extracted \skill{235B} and the \texttt{format\_only} specification are applied unchanged to Llama-3.1 and Gemma-3 models. $\Delta$ is versus each model's \none{} baseline; \textbf{bold} marks the larger standard-score gain per row.}
\label{tab:cross_family}
\footnotesize\setlength{\tabcolsep}{5pt}
\begin{tabular}{ll rrr rrr}
\toprule
 & & \multicolumn{3}{c}{Standard Scoring (\%)} & \multicolumn{3}{c}{$\Delta$ vs.\ None (pp)} \\
\cmidrule(lr){3-5} \cmidrule(lr){6-8}
Family & Model & None & \skill{235B} & \texttt{fmt\_only} & \skill{235B} & \texttt{fmt\_only} & \texttt{name\%} \\
\midrule
Qwen3      & 8B   & 58.0 & 69.3 & 79.3 & $+11.3$ & $\mathbf{+21.3}$ & 48\% \\
Qwen3      & 32B  & 77.7 & 66.3 & 69.7 & $-11.3$ & $\mathbf{-8.0}$ & 89\% \\
\midrule
Llama-3.1  & 8B   & 45.3 & 62.0 & 62.7 & $+16.7$ & $\mathbf{+17.3}$ & 37\% \\
Llama-3.1  & 70B  & 22.3 & 23.3 & 87.3 & $+1.0$ & $\mathbf{+65.0}$ & 1\% \\
\midrule
Gemma-3    & 4B   & 51.3 & 22.7 & 60.3 & $-28.7$ & $\mathbf{+9.0}$ & 63\% \\
Gemma-3    & 12B  & 38.3 & 22.3 & 70.0 & $-16.0$ & $\mathbf{+31.7}$ & 0\% \\
Gemma-3    & 27B  & 26.7 & 28.3 & 64.3 & $+1.7$ & $\mathbf{+37.7}$ & 16\% \\
\bottomrule
\end{tabular}
\par\vskip 2pt
\raggedright\footnotesize
\texttt{name\%} is the fraction of \none{} predictions using the evaluator-expected \texttt{name} JSON key. Format-only exceeds the full skill in all five non-Qwen rows and both Qwen anchor rows after all-call rescoring; Appendix~\ref{sec:appendix_crossfamily} explains the alias-boundary sensitivity.
\end{table*}

Alias-boundary sensitivity supports reading these rows as interface-alignment
evidence rather than portable procedural transfer. For
Qwen3-8B, accepting the current key aliases reduces the \skill{235B} delta from
$+11.3$ to $+3.7$~pp and the format-only delta from $+21.3$ to $+11.7$~pp. For
Llama-3.1-70B, the striking strict format-only gain ($+65.0$~pp) falls to
$+5.7$~pp after alias expansion, showing that most of the standard-score gain is
wrapper-contract alignment. Gemma rows are more model-dependent, but
format-only explains the pattern better than transfer of Qwen3-derived
procedures.

\subsection{API-Bank Results}
\label{sec:appendix_apibank}

The full API-Bank native matrix shows why cross-contract checks are necessary:
API-Bank rewards its own bracketed API-call contract, so BFCL-style JSON cues
can look harmful until predictions are deterministically transcoded back to the
native API-Bank surface. Weighted native API/required-parameter match is
50.2\% for \none{}, 50.7\% for the native API cue, only 7.5\% for the BFCL JSON
cue, and 51.0\% after prediction-only BFCL-to-API transcoding. The level-wise
rows follow the same pattern: BFCL JSON is near zero for most 1.7B/14B/32B rows
and low even for 8B, while deterministic transcoding restores native scores to
the no-skill/API-cue range. This check is therefore summarized here rather
than printed as another full matrix.

\subsection{API-Bank Target-Native Skill Boundary}
\label{sec:appendix_apibank_native}

Table~\ref{tab:apibank_native_skill} is the central API-Bank boundary table.
It uses a source-file-held-out split and compares the extracted API-Bank skill
with a length-matched generic procedural prompt under the native API/required
parameter metric.

\begin{table}[t]
\centering
\caption{API-Bank target-native skill boundary under source-file-held-out evaluation. The metric is native API/required-parameter match (\%). The generic row is a length-matched, hand-written procedural instruction, so the final delta is the conservative residual over the generic prompt. Interpretation: the weighted residual over the generic prompt is only +0.2~pp for 8B and +0.4~pp for 14B, so the target-native gain is almost entirely reproduced by generic procedural guidance.}
\label{tab:apibank_native_skill}
\scriptsize
\setlength{\tabcolsep}{2.3pt}
\begin{tabular}{llrrrrr}
\toprule
Model & Level & None & API cue & Generic & API skill & $\Delta$ skill--gen. \\
\midrule
8B & 1 & 50.0 & 53.8 & 76.2 & 75.6 & \ensuremath{-0.6} \\
8B & 2 & 55.7 & 57.2 & 78.9 & 79.4 & \ensuremath{+0.5} \\
8B & 3 & 45.0 & 43.0 & 43.0 & 44.0 & \ensuremath{+1.0} \\
\midrule
14B & 1 & 50.6 & 47.5 & 75.0 & 75.6 & \ensuremath{+0.6} \\
14B & 2 & 54.6 & 53.1 & 78.4 & 78.4 & \ensuremath{+0.0} \\
14B & 3 & 41.0 & 40.0 & 44.0 & 45.0 & \ensuremath{+1.0} \\
\midrule
8B & weighted & 51.3 & 52.9 & 70.0 & 70.3 & \ensuremath{+0.2} \\
14B & weighted & 50.2 & 48.2 & 69.6 & 70.0 & \ensuremath{+0.4} \\
\bottomrule
\end{tabular}
\end{table}

The near-parity does not depend on one generic wording: four generic procedural
variants remain within $-0.9$ to $+0.4$~pp of \skill{32B} under the native
metric. Paired native transitions are also small, at most one net
skill-favoring item per level for Qwen3-8B and Qwen3-14B. Stricter
parameter-set matching leaves generic variants closely paired with
\skill{32B}, with generic-minus-skill net transitions from $-4$ to $+2$ out of
454 held-out examples. Row-fresh local official-style execution likewise keeps
the best generic prompt within 0.6~pp on Qwen3-8B and 0.4~pp on Qwen3-14B, and
all paired intervals include zero. Unsupported-row execution, contract surgery,
non-Qwen transfer, and upstream-style prompt variants are boundary checks with
the same interpretation: the target-native API-Bank gain is real, but the
residual over generic procedural guidance is small, unstable, or
model-dependent rather than clear evidence of extracted procedural transfer.

\subsection{Prompt Position Ablation}
\label{sec:appendix_position}

We also test whether the format-scaffolding mechanism depends on where the
skill text is placed in the prompt. For 14B, the format-only advantage over full
skill is preserved in both slots: $+26.7$~pp in the system message and
$+33.0$~pp in the user prefix. For 8B, both placements preserve the advantage,
although the gap narrows from $+10.0$~pp in the system position to $+3.7$~pp in
the user prefix. The effect is therefore position-sensitive in magnitude but is
not an artifact of a single prompt slot.

\subsection{Threshold Sensitivity for Harmful-Cell Flip Rate}
\label{sec:appendix_threshold}

Varying the neutral-band threshold does not change the interpretation. With
$T=2$, $3$, and $5$~pp, the harmful-cell flip rates are 3/17 (18\%), 3/15
(20\%), and 3/13 (23\%), respectively. Thus key-alias canonicalization explains
an important subset of harms, notably for 14B, but it does not absorb all
interface- or prompt-induced failures.

\subsection{Error Type Breakdown}
\label{sec:appendix_error_breakdown}

We decompose BFCL errors for 1.7B, 8B, and 14B under \none{} and
\skill{235B}. ``Alias-only'' means standard scoring is wrong while
canonicalized scoring is correct. For 14B, skill injection increases
alias-only errors from 40 to 72; for 8B, it decreases alias-only errors from 29
to 6 and raises correct all-call answers from 174 to 208. This is consistent
with skill injection changing format compliance more directly than
function-calling competence.

As an additional local audit, we rescored the existing BFCL outputs with a
required-content diagnostic that requires the correct call count, function
names, and ground-truth parameter values while ignoring unsupported extra
predicted parameters. On the analyzed files, this diagnostic exactly matches
\textsc{bfcl-canonical}; no displayed main Qwen or cross-family condition has
an extra-argument-only residual. Thus \textsc{bfcl-canonical} is not gaining
credit by forgiving extra predicted parameters in this subset, although it
remains a local diagnostic rather than official execution.

We further audit the same outputs by explicitly decoding model strings into
function calls, then applying the upstream BFCL AST checker for callable
categories and the no-call rule for irrelevance. Across the 12 main Qwen rows,
local canonical and official-AST canonical scores differ by at most 1.7~pp, with
98.0--99.3\% item-level agreement. This supports using
\textsc{bfcl-canonical} as a Qwen-row diagnostic rather than a custom
value-checker artifact. It is not an official leaderboard rescore because call
decoding remains local, and it should not be generalized to all families:
Llama-3.1-70B has stricter type/value failures under the official AST checker,
with official-AST canonical scores 7.3--20.3~pp below local canonical scores.

\subsection{BFCL Official-Prompt Controls}
\label{sec:appendix_official_controls}

The official-prompt controls are boundary checks, not a second main-results
section. They ask whether the attribution survives under a BFCL
prompt contract closer to upstream usage. On a 300-row non-live Qwen subset,
Qwen3-8B and Qwen3-14B are near ceiling and \skill{235B} has no reliable paired
advantage over no-skill, official-format, or neutral prompts. On the full
1{,}040-row non-Qwen prompt-mode check, Llama-3.1-8B does not preserve the
Qwen-derived skill advantage: \skill{235B} is $0.9$~pp below no-skill and
$1.9$~pp below official-format, with the latter paired comparison significant.
Gemma-3-4B remains a model-specific boundary case, with \skill{235B} above
no-skill by $3.1$~pp and above official-format and neutral controls by $9.6$
and $7.9$~pp.

The stronger procedure-only control gives generic function-calling task-policy
advice without wrapper-key or JSON contract cues. Across Qwen3-1.7B, 8B, and
14B, it does not create a positive residual for the extracted skill:
procedure-only is $0.7$~pp below \skill{235B} on 1.7B, exactly tied with
\skill{235B} and no-skill on 8B, and $0.4$~pp below \skill{235B} on 14B; all
paired comparisons with \skill{235B} are non-significant. Additional
multi-neutral, prompt-variant, repaired-source, content-only, larger-Qwen, and
same-family Llama audits follow the same pattern. Once the upstream-style prompt
already states the output contract, any remaining skill effect is small,
model-dependent, recoverable by matched neutral wording, or matched by generic
procedure advice.

\subsection{Extraction Seed Variance}
\label{sec:appendix_seed_variance}

Across three balanced extraction seeds, the mean direction is negative for
1.7B, 8B, 32B, and 235B, while 14B shows high variance with a 31.7~pp spread.
This robustness check supports the repaired/balanced-induction interpretation:
the
original gains are not stable once the demonstration pool is controlled, and
the 14B sensitivity is consistent with a capability tipping point where small
changes in skill content produce large effects.

\subsection{Model Size and Compute Budget}
\label{sec:appendix_compute}

The open-weight evaluation models are Qwen3 0.6B, 1.7B, 4B, 8B, 14B,
32B, 30B-A3B, and 235B-A22B; Llama-3.1 8B/70B; and Gemma-3
4B/12B/27B. Original skills use GPT-4o for induction; because GPT-4o is
a closed API model, its parameter count and provider-side compute are not
publicly reported. We do not train or fine-tune any model. All open-weight
runs are inference-only, use temperature 0, and run on an internal NVIDIA H200
GPU cluster. The retained JSON traces contain 155{,}484 item-level generations
with 121.6 hours of summed per-item generation wall time. This is a trace-level
budget proxy rather than a scheduler-normalized GPU-hour total, because legacy
artifacts do not consistently record tensor-parallel allocation, queueing, or
scoring overhead.

\subsection{Reproducibility Checklist}
\label{sec:appendix_repro_checklist}

The release retains the data subsets, prompt/skill texts, raw generations,
score files, and analysis scripts needed to regenerate the reported diagnostics:
BFCL local and official-prompt audits, API-Bank native and row-fresh checks,
the Spider structured-output check, and the displayed tables and paired
statistics. Legacy GPT-4o extraction traces are treated as audited historical
inputs rather than token-replay targets because provider-side model parameters
and compute are not public.

\subsection{Claim Audit: Prose-Only Statistics}
\label{sec:appendix_audit}

Compact prose statistics are derived from the displayed result matrices and
retained analysis outputs: the format-shift regression uses the 8-point
Qwen3 BFCL-Contract key-shift analysis; harmful-cell flips use the full
standard-vs-canonicalized matrix; repaired and balanced effects use the
induction-subset tables; API-Bank transcode recovery and target-native residuals
use the source-file-held-out API-Bank diagnostics; and cross-family mean
absolute effects use the seven-model cross-family suite. These provenance links
are included to make the compact prose claims auditable without adding more
appendix tables.

\clearpage
\section{Item-Level Qualitative Case Studies}
\label{sec:case_studies}

This appendix keeps only representative item-level checks for the mechanisms
analyzed in Section~\ref{sec:attribution}. The aggregate evidence is in the
main text and Appendix~\ref{sec:appendix_fullmatrix_std}--\ref{sec:appendix_official_controls};
the examples below are included to make the error modes concrete rather than
to introduce additional claims.

\paragraph{Key-alias sensitivity.}
Qwen3-14B provides a direct example of scorer-side aliasing. Without an
injected skill, the model often emits BFCL-compatible calls using
\texttt{"name"} and a parameter object. After \skill{235B} is injected, many
otherwise identical calls switch to \texttt{"function\_name"}, a wrapper key
that the standard BFCL-style scorer rejects. On a \texttt{math.gcd} item, for
example, both predictions choose the same function and arguments
\texttt{\{"num1": 12, "num2": 15\}}; only the outer wrapper key changes. This
pattern is systematic: 35 of 48 standard-score regressions become correct under
canonicalized scoring, and 38 regressions involve a
\texttt{"name"}$\rightarrow$\texttt{"function\_name"} switch. The item-level
failure is therefore not a wrong procedure or wrong argument extraction, but a
benchmark-interface mismatch.

\paragraph{Format-only sufficiency.}
Qwen3-8B shows the complementary positive case. On one compound-interest item,
both \none{} and full \skill{235B} produce a bare JSON argument object with the
right values but no function-name wrapper, so both are marked wrong. The
\texttt{format\_only} prompt, which strips procedural SOP text and keeps only
the output wrapper template, produces the correct wrapped call. On a
portfolio-future-value item, both \skill{235B} and \texttt{format\_only} are
correct, differing only in whether the payload key is \texttt{"parameters"} or
\texttt{"arguments"}. Across all 300 local BFCL items, \texttt{format\_only} is
correct where \skill{235B} is wrong on 39 items, while the reverse happens on
only 9 items; aggregate accuracy is 58.0\% for \none{}, 69.3\% for
\skill{235B}, and 79.3\% for \texttt{format\_only}.

\paragraph{Original versus repaired skill.}
The repaired-skill comparison shows the same mechanism from the opposite
direction. The original \skill{235B} was induced from demonstrations that happen
to use BFCL-compatible \texttt{"name"}/\texttt{"arguments"} wrappers. The
repaired skill uses same-family Qwen3 demonstrations and no longer carries that
benchmark-aligned format cue. On a case-lookup item, the original skill changes
the 8B model from an incorrect \texttt{"function"} wrapper to the accepted
\texttt{"name"} wrapper; the repaired skill drops the wrapper again even though
the argument values remain semantically right. A prime-factor item has the same
shape. Across all 300 items, original \skill{235B} is correct where the
repaired skill is wrong on 62 items, while the reverse occurs on only 5 items
(12.4:1). This explains why original \skill{235B} reaches 69.3\% on 8B but
repaired \skill{235B} falls to 50.3\%, below the 58.0\% no-skill baseline.

\end{document}